\documentclass[twocolumn]{revtex4-1}
\usepackage{graphicx}
\usepackage{color}

\begin{document}

\title{Applying a new method to the 2-d Ising transition}
\author{Serge Galam}
\affiliation{National Center for Scientific research (CNRS), France}
\altaffiliation{ serge.galam@cnrs-bellevue.fr}
\author{Andr\'e C. R. Martins}
\affiliation{GRIFE -- EACH -- Universidade de S\~ao Paulo,\\
Rua Arlindo Bétio, 1000, 03828--000,  S\~ao Paulo, Brazil}
\altaffiliation{amartins@usp.br}

\begin{abstract}

The Ising ferromagnetic model on a square lattice is revisited using the Galam Unifying Frame (GUF), set to investigate the dynamics of two-state variable systems within the frame of opinion dynamics. When combined with Metropolis dynamics, an unexpected intermediate ``dis/order" phase is found  with the coexistence of two attractors associated respectively to an ordered and a disordered phases. The basin of attraction of initial conditions for the disordered phase attractor starts from zero size at a first critical temperature $T_{c1}$ to embody the total landscape of initial conditions at a second critical temperature $T_{c2}$ with $T_{c1}\approx 1.59$  and  $T_{c2}\approx 2.11$ in $J/k_B$ units. It appears that  $T_{c2}$ is close to the Onsager result $T_{c}\approx 2.27$.  The transition, which is first-order like, exhibits a vertical jump to the disorder phase at  $T_{c2}$, reminiscent of the rather abrupt vanishing of the corresponding Onsager second order transition. However, using Glauber dynamics  combined with GUF does not yield the intermediate phase and instead the expected single transition at $T_{c}\approx 3.09$. Accordingly,  although the "dis/order" phase produced by the GUF - Metropolis  combination is not physical, it is an intriguing result to be understood. In particular the fact that Glauber and Metropolis dynamics yield so different results using GUF needs an explanation. The possibility of extending GUF to larger clusters is discussed.

\end{abstract}

%Key words: Sociophysics, opinion dynamics, Ising spins, one dimensional chain

\maketitle

\section{Introduction}

The Ising model is a seminal model of statistical physics, which has inspired thousands of scientific papers \cite{binder, ising}. In particular the two-dimensional  nearest neighbor square lattice ferromagnetic Ising model is a cornerstone to the understanding of collective phenomena being the unique case to exhibit a second order phase transition with an exact analytical solution \cite{onsager}. All its properties are considered to be known.

Accordingly, in this work we use the two-dimensional  nearest neighbor ferromagnetic Ising model to investigate the properties of the Galam Unifying Frame (GUF), set to investigate the dynamics of two-state variable systems within the frame of opinion dynamics \cite{guf}. Applying GUF combined with Metropolis dynamics for the probability of a  single spin flip \cite{metropolis,glauber}, reveals an unexpected twofold order-disorder Ising transition. The transition is no longer at once from the ordered state into the disordered state but instead goes through an intermediate phase denoted ``dis/order" phase, for which two attractors exist, each one being associated to a different basin of attraction of initial conditions. Their respective sizes are a function of the temperature. 

The basin of initial conditions to the disordered phase starts from zero size at a first critical temperature $T_{c1}\approx 1.59$ to end up embodying the total landscape of initial conditions at a second critical temperature $T_{c2}\approx 2.11$. Temperature values are given in $J/k_B$ units where $J$ is the positive coupling constant and $k_B$ the Boltzmann constant. To make the presentation lighter from now on we take $J/k_B=1$.

For the Ising system, noting $p$ the proportion of spins in state $+1$ and $(1-p)$ the proportion of spins  in state $-1$, three different regimes are obtained for the dynamics as a function of temperature. 

(i) Starting at $T=0$ 
the phase diagram contains two attractors $p_-=0$ and $p_+=1$ with a separator $p_c=\frac{1}{2}$.
Increasing temperature moves the two attractors toward the separator with $p_->0$ and $p_+<1$. 

(ii) Then, at a temperature $T_{c1}$ the separator $p_c=\frac{1}{2}$ turns to an attractor. All these steps are expected from the usual description of the order/disorder Ising transition. However, while in the classical scheme, the change of status from separator to attractor is tuned on by the simultaneous merging of the two attractors  $p_-$ and $p_+$ at $p_c=\frac{1}{2}$, here the opposite process occurs. Two new symmetrical separators $p_{c-}<\frac{1}{2}$ and $p_{c+}=1-p_{c-}>\frac{1}{2}$ are expelled from $p_c=\frac{1}{2}$, which  becomes a third attractor. That is an unexpected process. 

(iii) Keeping on increasing the temperature, on each side of the attractor $p_c=\frac{1}{2}$, the corresponding attractor and the new separator move towards each other to eventually coalesce at a second critical temperature $T_{c2}$ and then disappear. In the range $T>T_{c2}$ only the attractor $p_c=\frac{1}{2}$ exists. 

At this stage it happens that  $T_{c2}\approx 2.11$ is close to the Onsager exact result $T_{c}\approx 2.27$.  Moreover, the transition, which is first-order like, exhibits a vertical jump to the disorder phase at  $T_{c2}$, reminiscent of the rather abrupt vanishing of the corresponding Onsager second order transition. 

However performing the same calculations as above using combining GUF with Glauber dynamics instead of Metropolis, restores the expected phase diagram of one single transition from the ordered phase into the disordered phase. The transition is continuous like and occurs at  $T_{c}\approx 3.09$ against $T_{c}=4$ for the Mean Field counterpart.

Accordingly,  although the "dis/order" phase produced by the GUF - Metropolis  combination is not physical, it is an intriguing result to be understood. In particular the fact that Glauber and Metropolis dynamics yield so different results using GUF needs an explanation. The possibility of extending GUF to larger clusters is also briefly discussed.

\section{Applying the GUF dynamics}

With no external field, the Hamiltonian for the  square lattice ferromagnetic Ising system is given by
\begin{equation}
H=-\frac{J}{2}  \sum_{\left[ i j \right]} \sigma_i \sigma_j,
\end{equation}
where $\left[ i j \right]$ a sum over all nearest neighbors $i$ and $j$ and $\sigma_i=\pm 1$. The associated magnetization is 
$M=\frac{1}{N}\sum_i^N \sigma_i=2p-1$ where $N$ is the total number of spins.

The problem has been solved  exactly with $T_c=2/\mathrm{arcsinh}(1)\approx 2.27$. In contrast, the classical mean field approach yields $T_c=4$. However in higher dimensions or with an external field, Monte Carlo (MC) simulations are required. They are implemented using a dynamics which respects detailed balance. The most common ones are  the Glauber and Metropolis dynamics \cite{metropolis,glauber}. 

In the Glauber dynamics, all spins are investigated in a sequential manner for each MC step. Given a spin $\sigma_i$ in a configuration $\mu_i$ with energy $E_{\mu_i}$, the flip  $\sigma_i\to -\sigma_i$ creates the new configuration $\eta_i$ with energy $E_{\eta_i}=-E_{\mu_i}$. The actual flip is implemented with the probability
\begin{equation}
\label{proba-g}
G_{\mu_i \to \eta_i}=\frac{1}{1+\exp(-2 E_{\mu_i}/k_BT)}.
\end{equation}

In Metropolis scheme, the spin is selected randomly and a MC step corresponds to N updates. The flip probability to the new configuration is given by
\begin{equation}
\label{proba-m}
M_{\mu_i \to \eta_i}=\min\{ 1, \exp [ 2 E_{\mu_i}/k_BT ] \}.
\end{equation}

In parallel a new scheme has been developed in the recent years to unify the large spectrum of discrete models proposed to study opinion dynamics. Indeed the Galam Unifying Scheme (GUF) was shown to embody all available discrete models of opinion dynamics \cite{guf}. We now apply it to the classical  ferromagnetic Ising model on a square lattice.

The GUF considers a polynomial development whose degree is given by the size of the group used to perform the update. Here 5 spins are used, the central one plus its 4 nearest neighbors yielding
\begin{equation}
\label{guf1}
p(t+1)=\sum_{k=0}^5 g_k [p(t)]^k [1-p(t)]^{5-k},
\end{equation}
where $p(t)$ is the proportion of $+1$ spins at time $t$ and $p(t+1)$ is the proportion of $+1$ spins at time $t+1$ after the equivalent of one MC step. The product $p(t)]^k [1-p(t)]^{5-k}$ is the probability to have a group of five
spins with $k$ spins $+1$ and $(5-k)$ spins $-1$, while  the coefficient $g_k$ is  the probability that the configuration ends up with the middle spin in the state $+1$ with the other 4 unchanged following either Glauber of Metropolis rule. By up-down symmetry $g_{0}=1-g_{5}$, $g_{1}=5 -g_{4}$ and $g_{2}= 10  -g_{3}$.
Performing the enumeration of the various configurations leads to respectively $g_5=1-b^2; g_4=5-4b; g_3=4$ and $g_5=1/(1+b^2); g_4=1/(1+b^2)+4/(1+b);  g_3=3+1/(1+b)$ for respective Metropolis and Glauber rules where temperature is incorporated using the parameter  $b\equiv e^{- 4/T}$. Plugging those coefficients into Eq. (\ref{guf1}) leads to respectively
\begin{eqnarray}
\label{m1}
p_M&=& p^5 ( 1-b^2) + p^4 (1-p)(5-4b)+4p^3 (1-p)^2 \nonumber\\
&+& 6 p^2 (1-p)^3 +4bp(1-p)^4 + b^2 (1-p)^5.
\end{eqnarray}
and
\begin{eqnarray}
\label{g1}
p_G&=& p^5 \frac{1}{1+b^2}  + p^4 (1-p)(\frac{1}{1+b^2}+\frac{4}{1+b})\\
&+& p^3 (1-p)^2 (3+\frac{4}{1 +b})+p^2 (1-p)^3 (7- \frac{4}{1+b})\nonumber\\
&+& p(1-p)^4 (5-\frac{1}{1+b^2}-\frac{4}{1+b}) + (1-p)^5 (1-\frac{1}{1+b^2} ), \nonumber
\end{eqnarray}
where $p=p(t)$ with $p_M=p(t+1)$ using the Metropoilis rule and $p_G=p(t+1)$ using the Glauber rule.

%%%%%%%%%%%%%%%%%%%%%%%%%%%%%%
\section{Uncovering a twofold transition}

To investigate the dynamics and phase transitions produced by Eqs. (\ref{m1}, \ref{g1}) we need to solve 
 the fixed point equation  $p(t+1)=p(t)$ for each case. The attractors correspond to the equilibrium states while the separator determines the flow direction when implementing the dynamical update.
 
 \subsection{GUF/Metropilis}

\begin{figure}
\includegraphics[width=.50\textwidth]{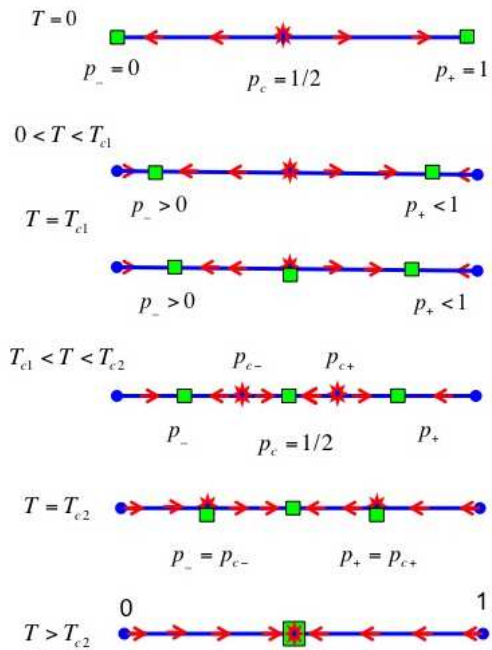}
\caption{Landscape of the attractors and separators as a function of temperature. 
The first top corresponds to $T=0$ with two attractors $p_=0$, $p_+=1$ and a separator $p_c=1/2$. Second from top corresponds to $0 \leq T < T_{c1}\approx 1.59$: same as before with $p_- \geq 0$ and $p_+\leq 1$. Third from top: at  $T=T_{c1}$ the separator is about to turn into an attractor giving rise to two symmetrical moving separators on each side. Third from bottom: for  $T_{c1} < T < T_{c2}$ three attractors $p_{-}$, $p_{+}$, $p_{c}$ separated by two the separators $p_{c-}$, $p_{c+}$. Second from bottom: at $T=T_{c2}$ the two separators  $p_{c-}$ and $p_{c+}$ coalesce with the two attractors  $p_{-}$, $p_{+}$ to suppress them. Bottom: only one attractor at  $p_c=1/2$ for $T > T_{c2}\approx 2.11$, which is close to the exact result $ 2.27$.}
\label{fpts}
\end{figure}

From $p_M=p$ we get always the solution $p_c=1/2$ with 3 different regimes. One with $p_c=1/2$ being the unique attractor. The second and third ones have respectively two and four symmetrical solutions up and bellow $p=1/2$ as shown in Figure \ref{fpts}. In the second case $p_c=1/2$  is a separator while it is an attractor in the third case. More precisely, above 
\begin{eqnarray}
 b_{c_2} &= &\frac{1}{5} \left(  
1-16(\frac{2}{3(5\sqrt{249}-9)})^(\frac{1}{3})+(\frac{2}{3})^{\frac{2}{3}}(5\sqrt{249}-9)^{\frac{1}{3}} \right) \nonumber \\
& \approx &0.150,
\end{eqnarray}
we observe only the zero-magnetization solution $p=1/2$. 

However, below  $b_{c_2}$, in the region defined by $0\leq b \leq b_{c_1} $ with
\begin{equation}
 b_{c_1} = \frac{1}{5}(\sqrt{41}-6)\approx 0.081,
\end{equation}
we have two solutions for $p$ given by
\begin{equation}
p_{\pm}=\frac{1}{2C}\left( C \pm \sqrt{C^2-4C( D- E)}
\right)
\label{pp}
\end{equation}
where $C\equiv (2b^2-8b+6)$, $D\equiv (3b^2-4b+1)$ and $E\equiv \sqrt{5b^4-8b^3+10b^2-8b+1}$. These solutions are attractors. They start from the values $p_-=0$ and $p_+=1$ at $b=0$ and move towards $p_c=1/2$ which behaves as a separator. However, they do not reach the zero magnetization as $b \rightarrow b_{c_2}$. Instead, they meet another pair of solutions 
\begin{equation}
p_{c\pm}=\frac{1}{2C}\left( C \pm \sqrt{C^2-4C( D+E)},
\right)
\label{ppp}
\end{equation}
that exist only in the interval
\begin{equation}
 b_{c_1}  \leq b \leq  b_{c_2}.
\end{equation}

As soon those two solutions $p_{c\pm}$ appear, they behave as separators and move towards $p_c=1/2$, which turns to an attractor as $b\rightarrow  b_{c_1}$. The  values $b_{c_1}$ and $b_{c_2} $ yield the critical temperatures of $T_{c_1}\approx 1.59$ and $T_{c_2}\approx 2.11$ using $T\equiv - 4/ \ln b$. 

We see that as soon as $b<b_{c_2}$, $p_c=1/2$ is no longer the only solution, leading us to believe that this is the actual phase transition. This provides a much better estimate of the critical temperature than that of mean field theory, since the correct value is $T_c \approx 2.27$. Notice that $T_{c_1}\approx 1.59$ had been previously observed for another attempt of estimating the critical temperature in the context of the GUF framework \cite{sousa}, however missing the existence of $T_{c_2}$.

The series of diagrams of  Figure \ref{fpts} shows the moving landscape of attractors and separators as a function of varying the temperature.  Two attractors $p_-=0$ and $p_+=1$ with a separator $p_c=\frac{1}{2}$ are found at $T=0$ as expected. Increasing temperature the two attractors move toward the separator with $p_->0$ and $p_+<1$ also as expected. However at a temperature $T_{c1}\approx 1.59$ the separator turns to an attractor $p_c=\frac{1}{2}$ with the simultaneous appearance of two new symmetrical separators $p_{c-}<\frac{1}{2}$ and $p_{c+}=1-p_{c-}>\frac{1}{2}$. That is an unexpected result. 

Increasing still the temperature, on each side of the attractor $p_c=\frac{1}{2}$, the attractor and the new separator move towards each other to eventually coalesce at $T_{c1}\approx 2.11$ and then disappear. At $T>T_{c2}$ only the attractor $p_c=\frac{1}{2}$ exists. 

\begin{figure}
\includegraphics[width=.50\textwidth]{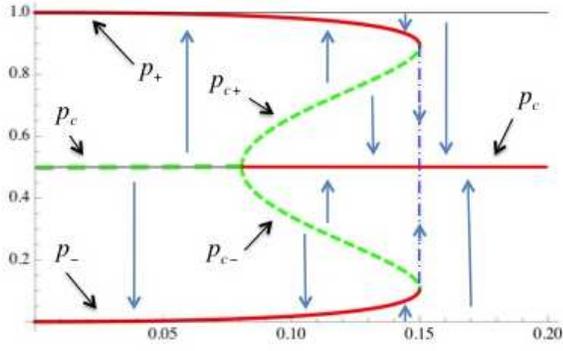}
\caption{The variation of all the five GUF fixed points as a function of $b=exp(-4/T)$ using Metropolis dynamics. The arrows show the flow direction while iterating the local updates.  Dark solid lines correspond to attractors, i.e., $p_{+}, p_{-}$ and $p_c$ for $b\geq b_{c1}$.  Dashed lines are separators, i.e.,  $p_{c+}, p_{c-}$ and $p_c$ for $b<b_{c1}$. }
\label{fpts-plus}
\end{figure}

Figure \ref{fpts-plus} shows the variation of all the five fixed points as a function of $b=\exp^{-\frac{4}{T}}$. The arrows shows the flow direction while iterating the local updates. Dark solid lines correspond to attractors, i.e., $p_{+}, p_{-}$ and $p_c$ for $b\geq b_{c1}$.  Dashed lines are separators, i.e.,  $p_{c+}, p_{c-}$ and $p_c$ for $b<b_{c1}$

\begin{figure}
\includegraphics[width=.20\textwidth]{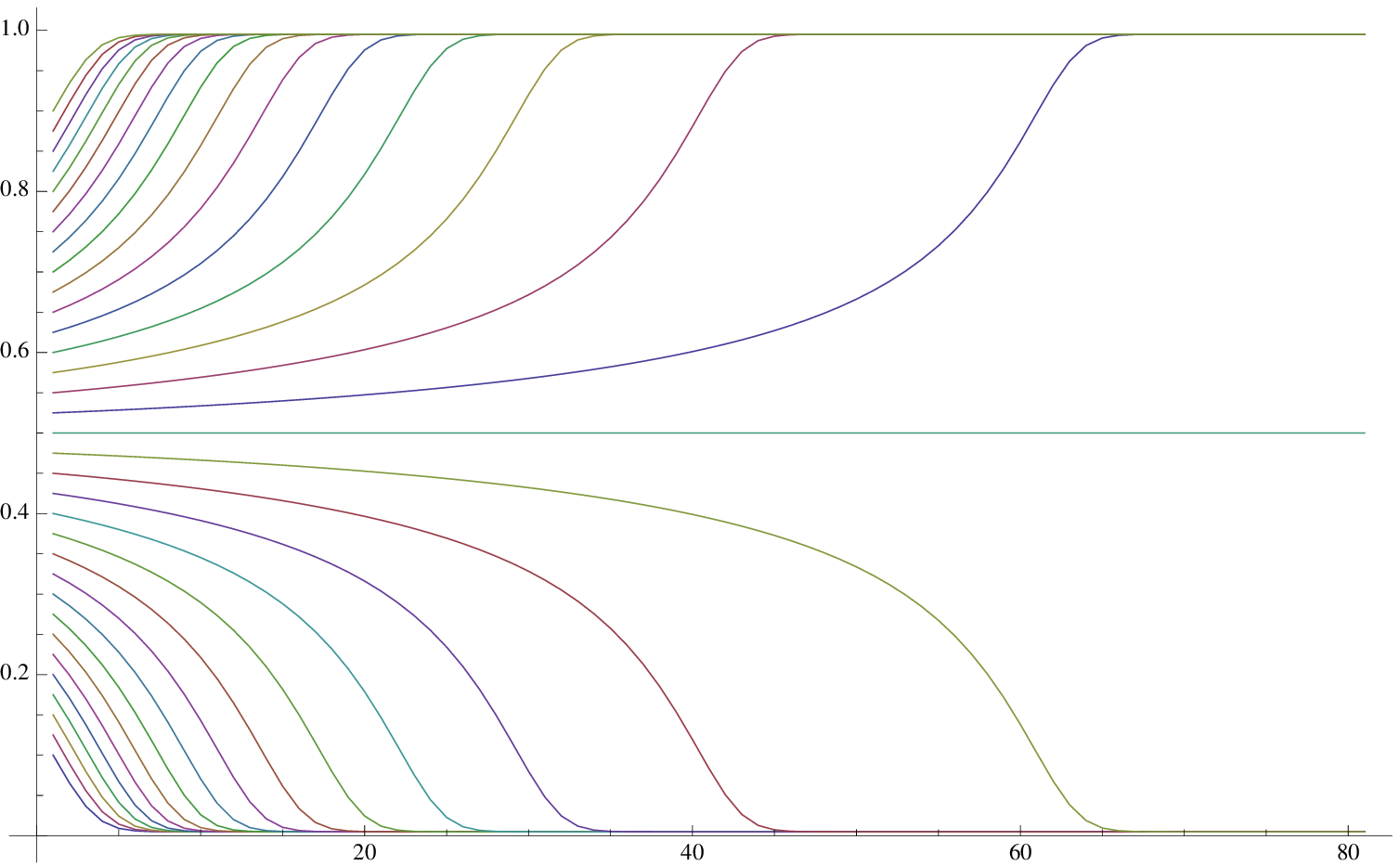}\quad
\includegraphics[width=.20\textwidth]{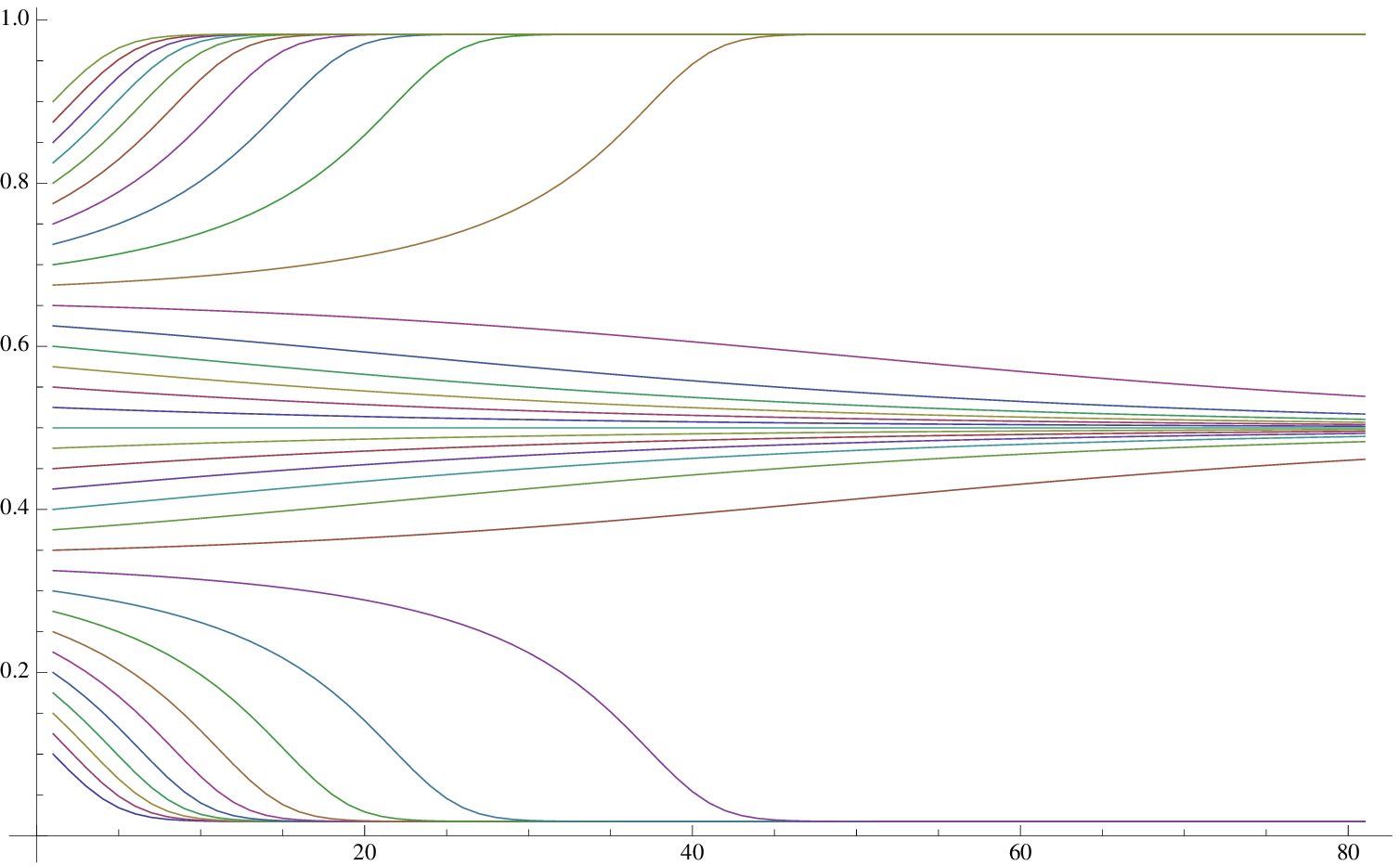}\\ 
\includegraphics[width=.20\textwidth]{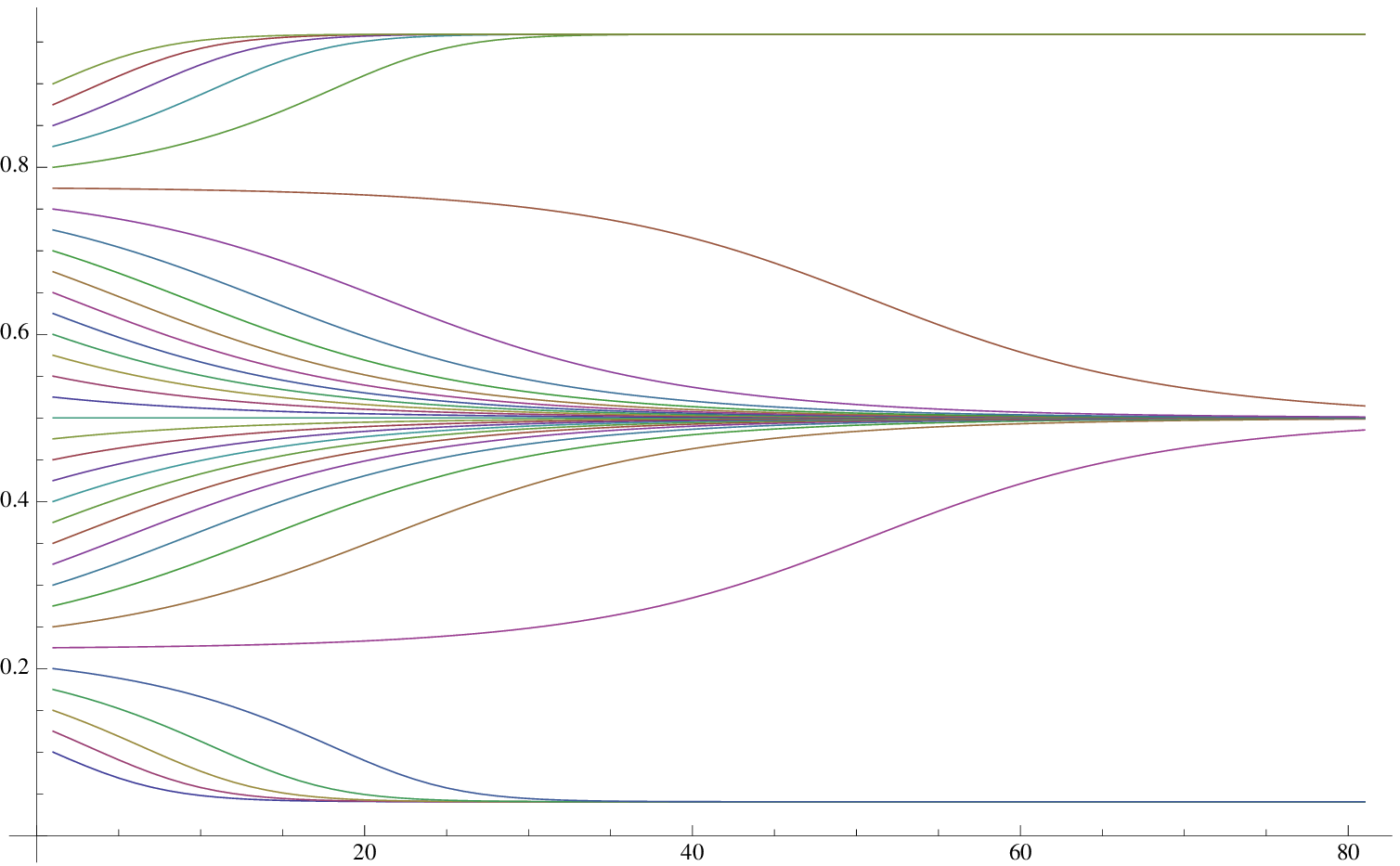}\quad
\includegraphics[width=.20\textwidth]{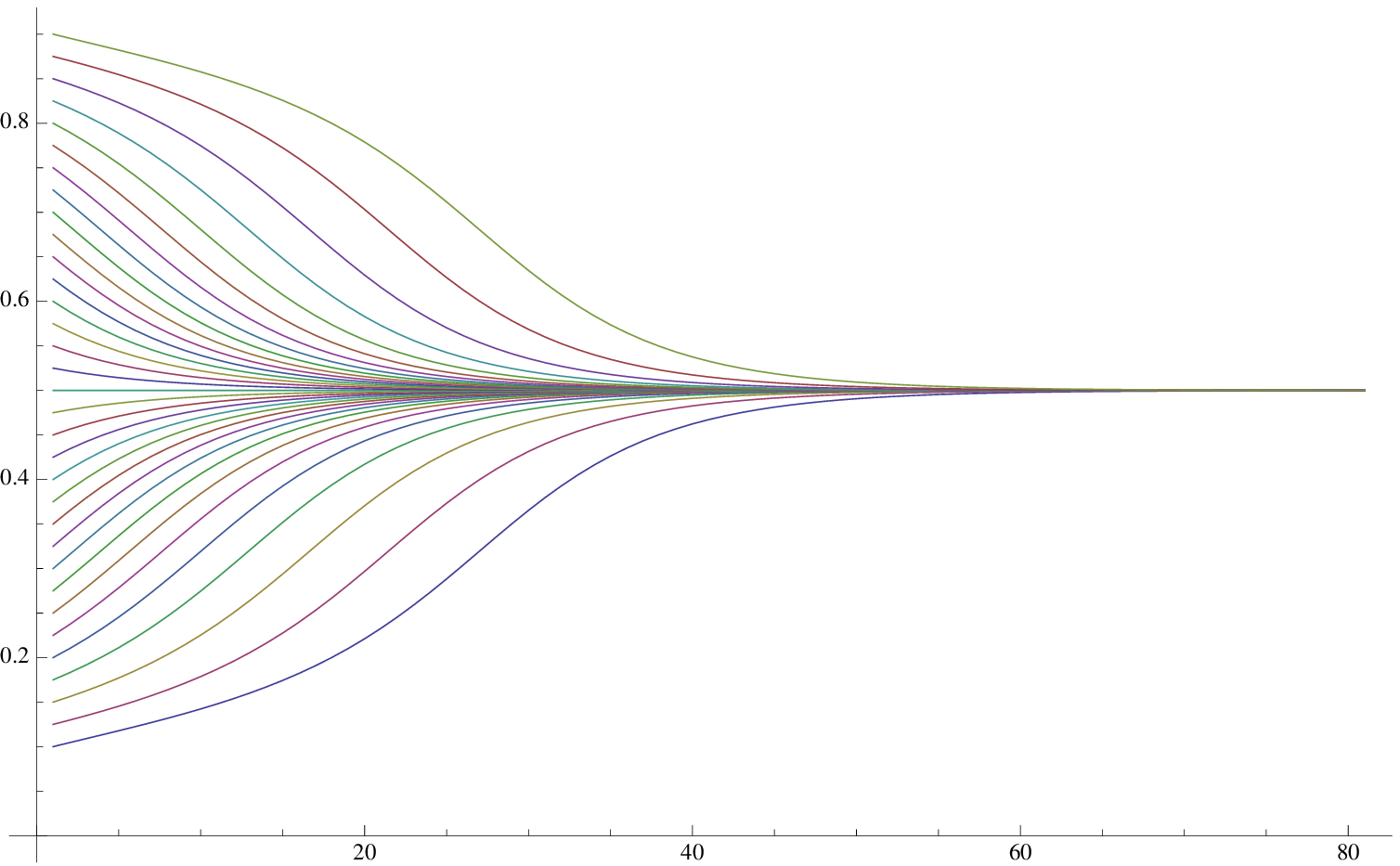}
\caption{The y axis shows the variation of $p_0$ under repeated application of the update rule $p_G$ using Eq. (\ref{m1}) from $p_0=010$ to $p_0=0.90$ with an incrustation of $+0.025$.  The x axis  shows the number of iterations up to 80. The series are shown for $b=0.006, 0.10, 0.13, 0.16$. Up left ($b=0.006$) corresponds to the usual ordered phase while bottom right ($b=0.16$) to the usual disordered phase.  Upright ($b=0.10$) and bottom left ($b=0.13$) show the uncovered new transition. The number of iterations are labeled along the abscise with 0, 20, 40, 60, 80. }
\label{ite}
\end{figure}

Figure \ref{ite} shows the three kind of dynamics obtained to describe the GUF  twofold Ising transition. At $b=0.006$ and $b=0.16$ stand the classical typologies with all initial configurations ending up at at an ordered phase with either a positive ($p_0>0.50$) or negative ($p_0<0.50$) magnetization for the first case and to zero magnetization for the second case. The Figures at $b=0.10$ and $b=0.13$ are unexpected with two possible equilibrium states depending on the initial configuration. It is worth to stress that the equilibrium state is not probabilistic. It is either $M\neq 0$ or $M=0$ depending on the value of $p_0$. 

\subsection{GUF/Glauber}

\begin{figure}
\includegraphics[width=.50\textwidth]{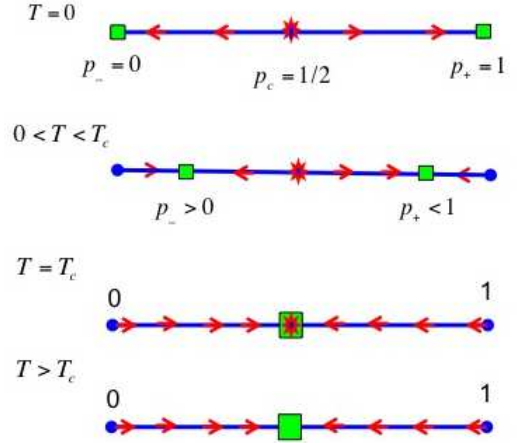}
\caption{Landscape of the attractors and separators as a function of temperature. 
The first top corresponds to $T=0$ with two attractors $p_=0$, $p_+=1$ and a separator $p_c=1/2$. Second from top corresponds to $0 \leq T < T_{c}\approx 3.09$: same as before with $p_- \geq 0$ and $p_+\leq 1$. Third from top: at  $T=T_{c}$,  $p_-$ and $p_+$ merge with the separator $p_c$, which in turn becomes an attractor. Bottom: only one attractor at $p_c=1/2$ for $T > T_c$ .}
\label{fpts-g}
\end{figure}

We now repeat above calculations using the Glauber scheme (Eq. \ref{proba-g}) instead of Metropolis (Eq. \ref{proba-m}). From Eq. \ref{g1} solving $p_G=p$  yields only three solutions, $p_c=1/2$ and 
\begin{equation}
p_{\pm}=\frac{1}{2S}\left(S\pm(-1+b)^{\frac{3}{2}}   \sqrt{T}\right) ,
\label{pp}
\end{equation}
were $S\equiv -1+3b-3b^2+b^3$ and $T\equiv -1+3b+b^2+5b^3$. The separators $p_{c\pm}$ obtained in the precedent Subsection do not exist any longer making the phase diagram shown in Figure \ref{fpts-g} very different from the one from Figure \ref{fpts}. However it is more like what is expected for the two-dimensional nearest neighbor ferromagnetic Ising model with no  "dis/order" phase and a single transition from the ordered phase into the disordered.

The associated critical temperature is obtained when $p_+=p_-=p_c=1/2$ yielding $b_c\approx 0.274$, which gives $T_c \approx 3.09$, whose value is in between the mean field result $T_c=4$ and the exact result $T_c \approx 2.27$.

\subsection{GUF/Glauber/Onsager/MF}

\begin{figure}
\includegraphics[width=.50\textwidth]{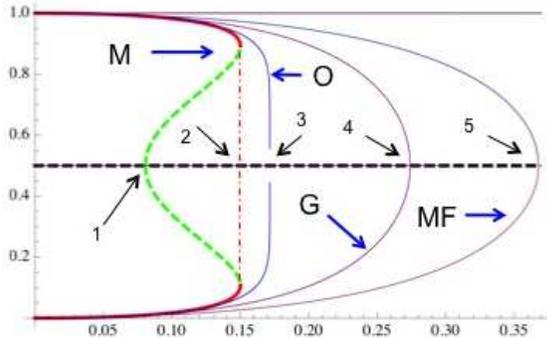}
\caption{Landscape of the attractors and separators for $p$ as a function of temperature via the variable $b=\exp(-4/T)$  in the abscise with Metropolis (M), Glauber (G), Onsager (O) and Mean Field (MF). Solid lines represents attractors while dashed lines are separators. However the line $p_c=1/2$ is a separator for M, G, O and MF only when $b<b_{c1}$ (1). For $b_{c1} (1) <b<b_{c2} (2)$, $p_c$ is an attractor for M and a separator for O, G and MF. However, for M other attractors coexist. When $b_{c2} (2)<b<b_c (3)$, $p_c$ is the unique attractor for M and still a separator for O, G and MF. For $b_c (3)<b< b_c(4)$, $p_c$ is an attractor for M, O, but still a separator for G, MF. When $b_c (4)<b< b_c(5)$, $p_c$ is an attractor for M, O, G, but still a separator for MF. Only when $b>b_c (5)$, $p_c=1/2$ is the unique attractor for M, O, G and MF. We have $b_{c1}\approx 0.081 , b_{c2}\approx 0.150, b_{c}(3)\approx 0.171, b_{c}(4)\approx 0.274, b_{c}(5)\approx 0.368$ yielding respectively $T_{c1}\approx 1.59, T_{c2}\approx 2.11, T_{c}(3)\approx 2.27, T_{c}(4)\approx 3.09, T_{c}(5)=4$.
 }
\label{m-g-o-mf}
\end{figure}

To compare the results obtained with respectively Metropolis and Glauber, the associated attractor and separator curves are exhibited  in Figure \ref{m-g-o-mf} as a function of temperature via the variable  $b$ in abscise together with the Onsager exact solution  for the magnetization \cite{yang} 
\begin{equation}
m_O=\left( 1-\sinh (2/T)^{-4}\right)^\frac{1}{8} ,
\end{equation}
where $2/T =-1/(2\ln b)$ and  the Mean Field formula
\begin{equation}
m_{MF}=\tanh \left(\frac{4m_{MF} }{T}\right) 
\end{equation}
using $p=(m+1)/2$ with M denoting Metropolis, G Glauber, O Onsager and MF Mean Field. Five critical values $b_c$ are obtained with (1, 2) for Metropolis, (3) for Onsager, (4) for Glauber and (5) for Mean Field.

From Figure \ref{m-g-o-mf} it is worth to notice that Metropolis yields a first-order like transition  with a critical temperature  $b_{c2}\approx 0.150$, which is close to Onsager exact result $b_{c}(3)\approx 0.171$, in addition to exhibit a vertical jump to the disorder phase at  $b_{c2}$, reminiscent of the rather abrupt vanishing of the corresponding Onsager second order transition. On the other hand, Glauber  is more like MF with an improvement in the value of the critical temperature at $b_{c}(4)\approx 0.274$ instead of  $b_{c}(5)\approx 0.368$ for Mean Field.

Solid lines represents attractors while dashed lines are separators. However, the line $p_c=1/2$ is a separator for M, G, O and MF only when $b<b_{c1}$ (1). For $b_{c1} (1) <b<b_{c2} (2)$, $p_c$ is an attractor for M and a separator for O, G and MF. But for M other attractors coexist. When $b_{c2} (2)<b<b_c (3)$, $p_c$ is the unique attractor for M and still a separator for O, G and MF. For $b_c (3)<b< b_c(4)$, $p_c$ is an attractor for M, O, but still a separator for G, MF. When $b_c (4)<b< b_c(5)$, $p_c$ is an attractor for M, O, G, but still a separator for MF. Only when $b>b_c (5)$, $p_c=1/2$ is the unique attractor for M, O, G and MF. We have $b_{c1}\approx 0.081 , b_{c2}\approx 0.150, b_{c}(3)\approx 0.171, b_{c}(4)\approx 0.274, b_{c}(5)\approx 0.368$ yielding respectively $T_{c1}\approx 1.59, T_{c2}\approx 2.11, T_{c}(3)\approx 2.27, T_{c}(4)\approx 3.09, T_{c}(5)=4$.

%%%%%%%%%%%%%%%%%%%%%%%

\subsection{GUF/Glauber bis}

\begin{figure}
\includegraphics[width=.50\textwidth]{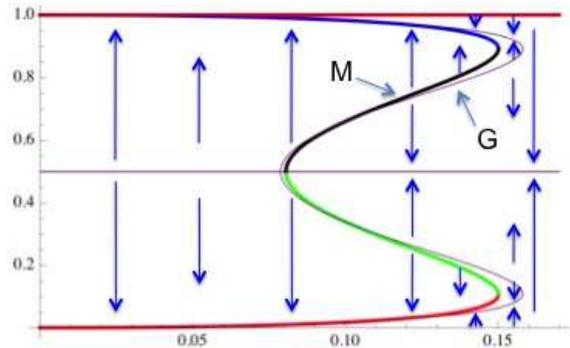}
\caption{The variation of all the five GUF fixed points as a function of $b=\exp^{-\frac{4}{T}}$. The arrows show the flow direction while iterating the local updates. G denotes the wrong Glauber scheme while M is for Metroplolis. }
\label{m-g-bis}
\end{figure}

In an earlier version of the paper, a misprint in Eq. (\ref{g1}) in the coefficient  $g_3=(3+\frac{4}{1 +b})$ of $p^3 (1-p)^2$ had been carried on during the iteration of the calculations previously done with Metropolis, leading to a very similar phase diagram as obtained with Metropolis  in Figure \ref{m-g-bis}. 

More precisely, putting $\frac{1}{1 +b}$ instead of $ \frac{4}{1 +b}$ in $g_3$, which implies the same change in $g_2=10-g_3$ recovers the twofold phase diagram one obtained using Metropolis with
\begin{eqnarray}
b_{c_1}&= &   \frac{1}{60} \nonumber \\
& & \left(-7 - \frac{61 11^{2/3}}{1147 + \sqrt{1059}))^{1/3}}
  + (11 (1147 + \sqrt{1059}))^{1/3}\right)\nonumber \\
 & \approx &0.079 ,
\end{eqnarray}

and $b_{c_2}$ is the solution of 
\begin{equation}
-1 - 6 b + 3 b^2 - 32 b^3 + 3 b^4 - 18 b^5 + b^6 =0,
\label{ppp}
\end{equation}
which cannot be solved analytically. However, the numerical solution is $b_{c_2}=0.158$. With respect to the four fixed points, we have
\begin{equation}
p_{\pm}=\frac{1}{2F}\left( F \pm \sqrt{F^2-4F( D- H)}
\right)
\label{pp}
\end{equation}
where $F\equiv (-6b^2-6)$, $G\equiv (b^3-3b^2+3b-1)$ and $H\equiv \sqrt{b^6-18b^5+3b^4-32b^3+3b^2-6b+1}$. 

As for Metropolis these solutions are attractors starting from the values $p_-=0$ and $p_+=1$ at $b=0$ and moving towards $p_c=1/2$ which behaves as a separator as $b$ is increased. At $b= b_{c_2}$ they coalesce with\begin{equation}
p_{c\pm}=\frac{1}{2F}\left( F \pm \sqrt{F^2-4F( G+H)} \right).
\label{ppp}
\end{equation}

The critical temperatures associated to $b_{c_1}$ and $b_{c_2} $ are  $T_{c_1}\approx 1.57$ and $T_{c_2}\approx 2.17$, which is rather close to the exact value $T_{c_2}\approx 2.27$. 

Although those results are wrong since resulting from a missing factor in Eq. (\ref{g1}), they are worth to be noticed due to the surprising identity with the ones obtained using Metropolis as illustrated in Figure \ref{m-g-bis}. Such a coincidence is worth more investigation. Indeed,  the finding of the ``dis/order" phase was unexpected using Metropolis thus prompting  the calculations to be carefully checked again and again from every part. However, the fact that Glauber was found to yield the same unexpected result as Metropolis, was expected, and thus made the misprint unnoticed in the coefficient $g_3$ in Eq. (\ref{g1}) with a rechecking of only the calculations starting from Eq. (\ref{g1}).

\section{Discussion}

In this work, the application of the GUF  combined with either Metropolis or Glauber schemes to the classical two-dimensional ferromagnetic  Ising model, was shown to exhibit some rather unexpected results. In particular the Metropolis scheme leads to the appearance of an intermediate  "dis/order" phase  between the ordered and disordered phases, turning first-order like the associated transition.  It happens that  the corresponding critical temperature $T_{c2}\approx 2.11$ is rather accurate with respect to the exact Onsager value $T_{c}\approx 2.27$. In addition, the transition exhibits a vertical jump to the disorder phase reminiscent of the rather abrupt vanishing of the corresponding Onsager second order transition. 

Accordingly,  although the "dis/order" phase produced by the GUF - Metropolis  combination is not physical, it is an intriguing result worth to be understood. 

In parallel, contrary to what could have been expected, combining Glauber dynamics to GUF restores the Ising single transition at $T_{c}\approx 3.09$, which arises the question to determine why Glauber and Metropolis dynamics lead to different equilibrium sates when combined with GUF in the case of the 2-d Ising model?

The various behaviors obtained respectively with GUF - Metropolis, GUF - Glauber, Onsager and Mean Field are shown together in Figure \ref{m-g-o-mf}, which provides some coherent picture about the effect of various approximations on departing from the exact treatment of the 2-d Ising model. At this stage, GUF needs more investigation to understand its nature and find the physical origins of those discrepancies with both the classical Mean Field and the exact result. The question has been evoked in a series of works  \cite{sousa, slani,lambiotteredner08a, exit,przybylaetal11a}. 

One promising direction is certainly to extend the cluster size to which GUF has been applied. Instead of a cluster of 5 spins limited to the nearest neighbors, it should be interesting to consider both, the inclusion of next-nearest neighbors with a 9 spins cluster and also the extension to 3-d with a 7 spins cluster, noticing that the respective topologies are instrumental in the calculation of the various GUF coefficients.

%%%%%%%%%%%%%%%%%%%%%

\end{document}